\documentclass[%
 reprint,
 amsmath,amssymb,
 aps,
showkeys,
citeautoscript]{revtex4-2}

\setcitestyle{super}

\newcommand*{\citen}[1]{%
  \begingroup
    \romannumeral-`\x 
    \setcitestyle{numbers}%
    \cite{#1}%
  \endgroup   
}

\usepackage{comment}
\usepackage{graphicx}
\graphicspath{{fig_maintext/}}

\usepackage{dcolumn}
\usepackage{bm}
\usepackage[colorlinks=true,bookmarks=false,citecolor=black,urlcolor=blue,linkcolor=black]{hyperref} 
\usepackage{amsmath}
\DeclareMathOperator*{\argmax}{arg\,max}

\usepackage{xcolor}
\usepackage{soul}
\soulregister\cite{7} 
\soulregister\ref{7} 
\soulregister\eqref{7} 

\begin{document}

\title{Adaptive Optical Multi-Spectral Matrix Approach for Label-free High-resolution Imaging through Complex Scattering Media}

\author{Yiwen Zhang$^{1,\dagger}$}
 \email{yzhang67@usc.edu}
\author{Minh Dinh$^{1,\dagger}$}
\author{Zeyu Wang$^{1}$}
\author{Tianhao Zhang$^{1}$}
\author{Tianhang Chen$^{1,2}$}
\author{Chia Wei Hsu$^1$}
\affiliation{
$^1$Ming Hsieh Department of Electrical and Computer Engineering, University of Southern California, Los Angeles, California 90089, USA
$^2$Interdisciplinary Center for Quantum Information, State Key Laboratory of Modern Optical Instrumentation, College of Information Science and Electronic Engineering, Zhejiang University, 310027 Hangzhou, China
$^\dagger$These authors contributed equally to this work\vspace{-3pt}
}

\begin{abstract} 
Imaging through complex scattering media is severely limited by aberrations and scattering which obscure images and reduce resolution.
Confocal and temporal gatings partly filter out multiple scattering but are severely degraded by wavefront distortions. Adaptive optics restore  resolution by correcting low-order aberrations and matrix-based imaging enables more complex wavefront corrections. However, they struggle to undo high-order aberrations under strong scattering, preventing imaging at greater depths.
To address these challenges, we present Scattering Matrix Tomography (SMT), an approach that makes full use of the wavefront engineering capability of scattering matrix and extreme adaptive optics.
SMT reformulates imaging through complex media as a numerical optimization and employs Zernike-mode wavefront regularization and coarse-to-fine nonconvex optimization strategy to reverse severe aberrations, enabling noninvasive high-resolution volumetric imaging in multiple scattering regime.
Based on the spectrally-resolved matrix measurement, SMT achieves a depth-over-resolution ratio above 900 beneath $ex~vivo$ mouse brain tissue and volumetric imaging at over three transport mean free paths inside an opaque colloid, where conventional methods fail to correct strong aberrations under these challenging conditions.
SMT is noninvasive, label-free, and works both inside and outside the scattering media, making it suitable for various applications, including medical imaging, biological science, device inspection, and colloidal physics.
\end{abstract}

\keywords{light scattering, computational aberration correction, computational imaging, scattering matrix tomography.\vspace{-3pt}}

\maketitle


\section{Introduction \label{sec: intro}}
\vspace{-9pt}

Optical imaging plays a crucial role in non-destructive, high-resolution visualization, owing to its high spatial resolution. 
It has been widely employed to study biological and physical processes, enabling applications such as cellular-level neuronal activity monitoring, diagnostic imaging in ophthalmology and audiology, and surgery-free biopsies. Beyond biology, optical imaging also holds great promise for applications like optical characterization of bulk colloids and high-resolution, non-destructive device testing.
However, optical imaging is significantly limited by its inability to penetrate deeply into complex scattering media, such as biological tissue, where inhomogeneities lead to aberrations and multiple scattering that degrade image quality even to the point of being unrecognizable ~\cite{2017_Badon_OE}.
While aberrations distort the light wavefront and reduce spatial resolving power, severely undermining the image contrast, multiple scattering exponentially attenuates the signal and introduces a speckled background, obscuring the already-aberrated desired information~\cite{yoon2020deep,bertolotti2022imaging}. 
Overcoming these challenges to achieve label-free, high-resolution imaging through complex scattering media is essential to expand the utility of optical imaging in practical applications.

Over the years, gating techniques have been developed to enhance image contrast, resolution, and depth in optical imaging by selectively isolating the single-scattered light from the unwanted background.
For example, the reflectance confocal microscopy (RCM)~\cite{calzavara2008reflectance, 2018_Xia_BOE} 
uses spatial gating. 
Optical coherence tomography (OCT)~\cite{fercher2003oct} and optical coherence microscopy (OCM)~\cite{aguirre2015ocm} apply a coherence time gate in addition. 
Other approaches, like selective detection of the forward scattered photons~\cite{2017_Zhao_OL, 2022_Cua_BOE} and photoacoustic tomography~\cite{2012_Wang_Science_review} image deeper with a reduced resolution.
Despite these efforts, these techniques are generally limited to a depth-over-resolution ratio of around or below 200 in biological tissue~\cite{2012_Wang_Science_review,2016_Lai_MICC_review,2017_Gigan_nphoton_feature} due to the inherent exponential decay of single-scattered light, meaning the high resolution is confined to superficial layers.
To further complicate matters, the single-scattered wave is affected by the unavoidable aberrations induced by the sample's heterogeneity, severely degrading image quality.

Adaptive optics (AO) has been a significant advancement in improving the quality of the spatial gate by correcting for low-order wavefront distortions~\cite{2010_Ji_nmeth,2021_Hampson_NRMP_review}
within a volume called the isoplanatic patch~\cite{2021_Hampson_NRMP_review,2017_Park_nmeth}. Traditional AO systems rely on guidestars or wavefront sensors to correct aberrations, while the image-based sensorless AO~\cite{2007_Debarre} using Zernike-mode based image metric optimization eliminates this need but still uses hardware with limited correction modes and speed. Computational AO~\cite{2011_Tippie_OE,2016_Hillmann_srep,adie2012computational,2015_Shemonski_nphoton} 
offers potentially more correction modes, faster speed, and better multiple-scattering immunity.
Despite these advancements, AO falls short of achieving high-resolution imaging in the presence of strong high-order aberrations and scattering. Higher-order aberrations require correcting both the incident and the outgoing wavefronts~\cite{2021_Liu_BOE}, but existing computational AO schemes can only perform one correction.
Furthermore, the severe multiple scattering 
causes large errors in AO, leading to a breakdown in performance and limiting its applicability in deeply scattering media.

{Recently, the matrix-based approach~\cite{kang2015imaging,badon2016smart} has gained significant attention for its potential to push the depth limits of optical imaging using the full sample's optical response.
Unlike confocal detection, the matrix-based approach captures signals arriving at both confocal and non-confocal positions, building a more complete picture of the sample’s input-output responses, known as the scattering matrix~\cite{popoff2010measuring, 2016_Mounaix_PRL}, which allows for the digital correction of both the incident and reflected wavefronts.}
Pioneering works on the ``closed-loop accumulation of single-scattering'' (CLASS) method~\cite{kang2017high,kim2019label,yoon2020laser,2023_Kwon_ncomms,2023_Li_OE}
and the ``distortion matrix'' approach~\cite{badon2020distortion,2024_Najar_NatComm,2024_Balondrade_NatPhoton} (whose latest version is equivalent to CLASS~\cite{2023_Bureau_ncomms}) have leveraged correlations within the reflection matrix, the reflection mode of the scattering matrix, to correct wavefront distortions.
Extending to the multi-spectral matrix imaging scheme,~\cite{2023_Lee_ncomms,2024_Balondrade_NatPhoton} 
the recent ``volumetric reflection-matrix microscopy'' (VRM)~\cite{2023_Lee_ncomms} generalizes CLASS to add dispersion compensation.
However, current matrix-based methods still face significant challenges when {the single scattering signal is severely distorted by high-order sample-induced aberrations and overwhelmed by the multiple scattering backgrounds, which introduce significant errors in the wavefront correction. 
To fully exploit the wealth of information contained in the scattering matrix, robust and effective wavefront correction approaches are required — ones that can withstand the detrimental effect of the strong multiple scattering to undo aberrations
and unlock more potential of matrix-based imaging for high-resolution imaging through complex media.
}

\begin{figure*}
    \includegraphics[scale = 0.87] {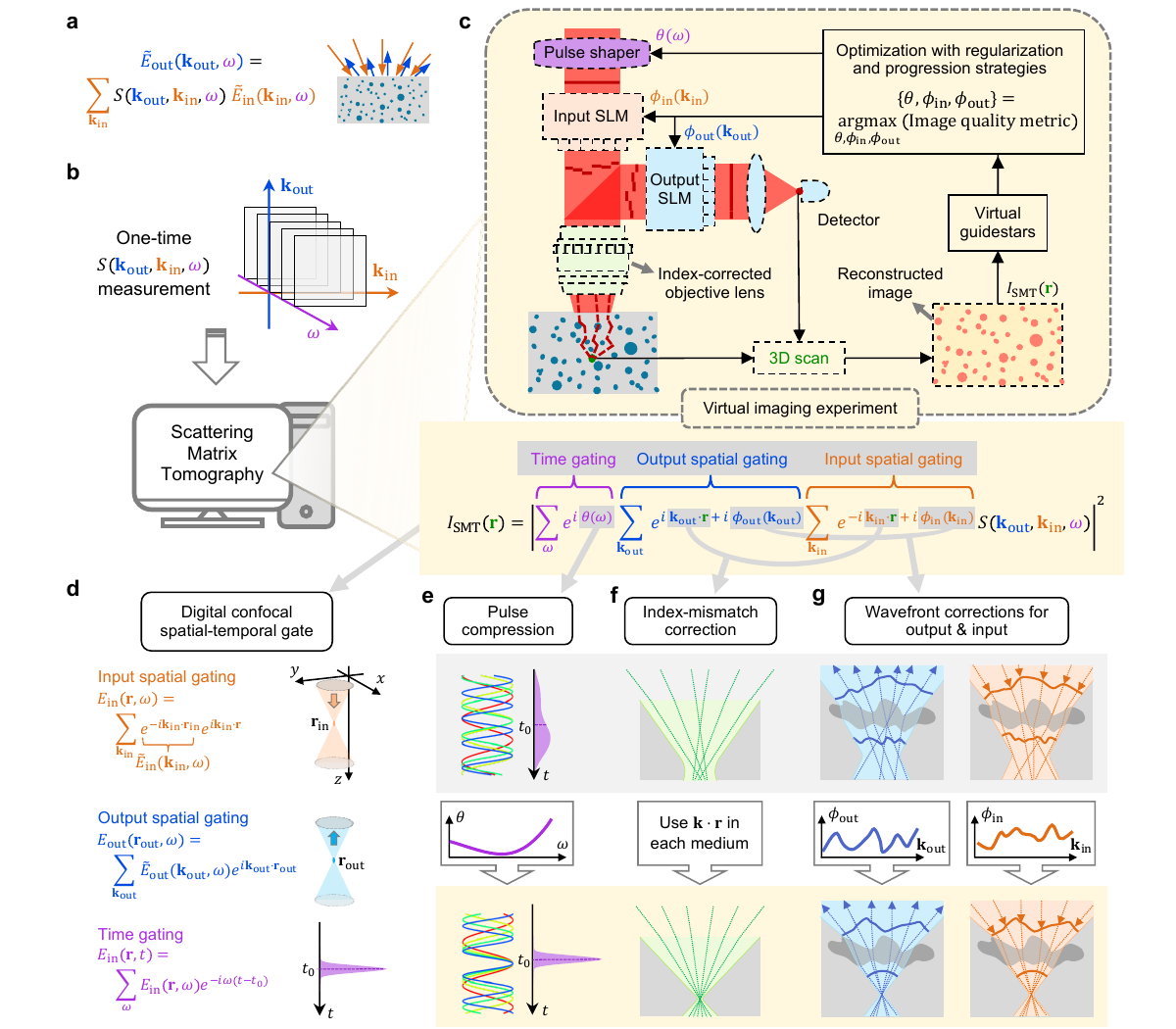}
    \caption{\textbf{Virtual imaging experiment and scattering matrix tomography (SMT).} 
    \textbf{a}, The scattering matrix $S(\bf k_{\rm out}, \bf k_{\rm in},\omega)$ relates any 
    incident field $\tilde{E}_{\rm in}(\bf k_{\rm in},\omega)$ to the resulting scattered field $\tilde{E}_{\rm out}(\bf k_{\rm out},\omega)$.
    \textbf{b}, After one-time measurements of each spectrally-resolved scattering matrix $S(\bf k_{\rm out}, \bf k_{\rm in},\omega)$, the data is processed computationally to reconstruct the image. 
    \textbf{c}, SMT imaging that digitally performs triple gating, spatiotemporal wavefront corrections, volumetric scanning, and optimization via Eqs.~\eqref{eq:smt_corrected}--\eqref{eq:opt}. The reconstructed image acts as virtual guidestars intrinsic to the sample, providing feedback noninvasively.
    \textbf{d}, The scattering matrix can synthesize input spatial gating, output spatial gating, and time gating by summing over the incident momentum ${\bf k}_{\rm in}$, outgoing momentum ${\bf k}_{\rm out}$, and frequency $\omega$ respectively.
    \textbf{e--g}, Dispersion, refractive index mismatch at interfaces, and wavefront distortions (from both the optical system and the sample, including both aberrations and multiple scattering) degrade the gates. SMT corrects all of them digitally through a frequency-dependent phase $\theta(\omega)$ that acts as a virtual pulse shaper, appropriate momenta and phase coefficients for the medium that act as a virtual index-corrected objective lens, and angle-dependent phases $\phi_{\rm in}(\bf k_{\rm in})$ and $\phi_{\rm out}(\bf k_{\rm out})$ that act as two virtual spatial light modulators (SLMs).
    }
    \label{fig1:smt} 
\end{figure*}

Here, we propose an adaptive optical multi-spectral matrix approach to tackle the challenge of achieving high-resolution, label-free tomographic imaging through complex scattering media.
{Relying on the sample’s full far-field linear response encapsulated in the scattering matrix, the approach, termed Scattering Matrix Tomography (SMT), serves as a unified platform for systematically exploring the rich information embedded in the scattering matrix, allowing us to perform scattering-based imaging with various gating mechanisms and apply flexible corrections.}
Furthermore, {inspired by adaptive optics, we formulate the problem of imaging through complex scattering media as a numerical optimization} rather than relying on matrix correlations. {To overcome the significant challenges posed by high-dimensionality and nonconvexity due to severe sample-induced aberrations and optimization error due to overwhelming multiple scattering, we employ wavefront regularizations and coarse-to-fine progressive optimization strategy along with additional refinements that further enhance the robustness and image quality.}
Together, these advancements enable our method to achieve a depth-over-resolution ratio of 910 when imaging a resolution target at one millimeter beneath {\it ex vivo} mouse brain tissue---the highest reported in the literature (including both optical and non-optical label-free methods) to our knowledge---where the signal is reduced by over ten-million-fold due to multiple scattering and is completely overwhelmed by speckles prior to the digital corrections.
We maintain the ideal transverse and axial resolutions across a depth of field of over 70 times the Rayleigh range at depths beyond three transport mean free paths inside an opaque colloid---the deepest reported to our knowledge.
We synthesize conventional reflection-based methods (RCM, OCT, and OCM) and the most advanced matrix-based method VRM, and found all of them to fail at these depths.
{By framing the challenges of imaging through scattering media as numerical reconstruction and optimization problems, employing wavefront regularization and the progressive optimization strategy, our method builds upon the strengths of adaptive optics and the scattering matrix while providing important advancements that enable deeper, higher-resolution imaging in complex scattering environments, offering an intuitive and versatile platform that uses wave physics and computation to push the capabilities of imaging.}

\section{Scattering Matrix Tomography \label{sec: smt}}
\subsection{Image Reconstruction Framework \label{sec: sub_smt}}

{The scattering matrix $S(\bf k_{\rm out}, \bf k_{\rm in},\omega)$ 
encapsulates the sample's far-field linear response~\cite{popoff2010measuring, 2016_Mounaix_PRL} (Fig.~\ref{fig1:smt}a). At frequency $\omega$, for each incidence with momentum $\bf k_{\rm in}$, the scattered fields are measured and stored at a corresponding matrix column, with each row corresponding to one output momentum $\bf k_{\rm out}$. Since any wavefront can be synthesized from plane waves with designed phases and amplitudes, the scattering matrix, with its flexible wavefront engineering capability, allows one to mimic imaging experiments with any arbitrary incident and scattered waves by computationally tailoring the matrix columns and rows, thus achieving desirable images.

In light of this, as illustrated in Fig.~\ref{fig1:smt}b-c, after a one-time measurement of a subset of $S(\bf k_{\rm out}, \bf k_{\rm in},\omega)$, we utilize this data to perform virtual imaging experiments --- 
digitally synthesizing the would-be response of the sample for any customized spatiotemporal input and any tailored measurement in space-time without additional physical experiments. 
In these virtual experiments, we can enforce all gating mechanisms (time gating and confocal spatial gating), all possible correction schemes (pulsing shaping, index-mismatch correction, and double-path corrections of high-order aberrations), scan the focus, and optimize the image quality, all carried out digitally without having to wait for the hardware}. Below, we provide a rigorous framework to reconstruct images and correct for wavefront distortions using the scattering matrix.

Any incident wave is a superposition of plane waves: ${\it E}_{\rm in}( {\bf r}, t) = \sum_{{\bf k}_{\rm in},\omega} \tilde{\it E}_{\rm in}({\bf k}_{\rm in},\omega) e^{i {\bf k}_{\rm in} \cdot {\bf r} - i\omega t}$, where ${\bf r} = (x,y,z)$ is the position, $t$ is time, and $\tilde{\it E}_{\rm in}({\bf k}_{\rm in},\omega)$ is the amplitude of its plane-wave component with incoming momentum $\bf k_{\rm in}$ at frequency $\omega$.
The amplitudes $\tilde{E}_{\rm out}({\bf k}_{\rm out}, \omega)$ that form the resulting outgoing wave 
${\it E}_{\rm out}( {\bf r}, t) = \sum_{{\bf k}_{\rm out},\omega} \tilde{\it E}_{\rm out}({\bf k}_{\rm out},\omega) e^{i {\bf k}_{\rm out} \cdot {\bf r} - i\omega t}$ 
are given by the scattering matrix through $\tilde{E}_{\rm out}({\bf k}_{\rm out},\omega) =\sum_{{\bf k}_{\rm in}} S({\bf k}_{\rm out}, {\bf k}_{\rm in},\omega)\tilde{E}_{\rm in}(\bf k_{\rm in},\omega)$.
The angular summations are restricted to $|{\bf k}|^2 = |{\bf k}_{\parallel}|^2 + k_z^2 = (n_{\rm bg} \omega/c)^2$ in a background medium with speed of light $c/n_{\rm bg}$; we use ``angle'' interchangeably with ``momentum.''

We perform the digital gatings through summations (Fig.~\ref{fig1:smt}d): summing over plane waves $e^{i{\bf k_{\rm in}}\cdot{\bf r}}$ with incident angle $\bf k_{\rm in}$ and amplitude $\tilde{E}_{\rm in}({\bf k}_{\rm in},\omega) = e^{-i{\bf k_{\rm in}}\cdot{\bf r_{\rm in}}}$ creates an incident beam ${\it E}_{\rm in}( {\bf r},\omega)$ spatially focused at $\bf r_{\rm in}$, forming an input spatial gate;
summing over plane waves $e^{i{\bf k_{\rm out}}\cdot{\bf r_{\rm out}}}$ with outgoing angle $\bf k_{\rm out}$ and amplitude $\tilde{E}_{\rm out}({\bf k}_{\rm out},\omega)$ from the scattering matrix yields the scattered field ${\it E}_{\rm out}( {\bf r} = {\bf r}_{\rm out},\omega)$ given a virtual detector at a position conjugated to ${\bf r}_{\rm out}$, forming an output spatial gate;
summing over frequencies $\omega$ with an $e^{-i\omega(t-t_0)}$ phase creates an incident pulse that arrives at $\bf r_{\rm in}$ at time $t=t_0$, forming a temporal gate.
Given a spatiotemporal focus arriving at $\bf r_{\rm in}$ at time $t_0$, the scattered field at position ${\bf r}_{\rm out}$ at time $t$ is therefore
$E_{\rm out}({\bf r}_{\rm out}, \it t) = \sum_{\bf k_{\rm out}, \bf k_{\rm in},\omega} S({\bf k}_{\rm out}, {\bf k}_{\rm in},\omega) e^{i {\bf k}_{\rm out} \cdot {\bf r}_{\rm out} - i {\bf k}_{\rm in}\cdot {\bf r}_{\rm in} - i \omega (t-t_0)}$.
In our virtual experiment, we align the input spatial focus with the output spatial detection (setting ${\bf r}_{\rm in} = {\bf r}_{\rm out} = {\bf r}$) and align the temporal focus with the confocal spatial one (evaluating $E_{\rm out}$ at time $t=t_0$), which yields
the triply-gated scattering amplitude of the sample at position ${\bf r}$, denoted as
\begin{equation}
    \psi_{\rm SMT}(\bf r) = \sum_{\bf k_{\rm out}, \bf k_{\rm in},\omega}
    \it e^{\it i(\bf k_{\rm out}-\bf k_{\rm in})\cdot{\bf r}}
    \it S(\bf k_{\rm out}, \bf k_{\rm in},\omega).
    \label{eq: psi_smt}
\end{equation}
Digitally scanning $\bf{r}$ (Fig.~\ref{fig1:smt}c) forms a phase-resolved image of the sample where the input spatial gate, output spatial gate, and temporal gate all align at every point in the absence of aberrations and scattering. 
The lateral spread of $\bf k_{\rm out}-\bf k_{\rm in}$ in the summation (typically set by the numerical aperture NA) determines the lateral resolution, with no degradation away from any particular focal plane. The axial spread (typically set by the spectral bandwidth) determines the axial resolution.
Like digital holography~\cite{2005_Ferraro_OE}, the $\psi_{\rm SMT}(\bf r)$ image can be volumetric or a slice with any orientation.
The triple summation averages away random noises, providing a signal-to-noise ratio advantage akin to that of frequency-domain OCT over time-domain OCT~\cite{2017_deBoer_OE_review}.
One can utilize any subset of the scattering matrix---reflection, transmission, remission~\cite{2017_Zhao_OL,2022_Bender_PNAS}, a combination of them, or other subsets---with any number of angles and frequencies. 
We use the non-uniform fast Fourier transform~\cite{barnett2019parallel} to efficiently evaluate these summations and the 3D spatial scan.
Eq.~\eqref{eq: psi_smt} is the minimal version of ``scattering matrix tomography'' (SMT).

{The triple gating suppresses multiple scattering.
The single-scattering field at output ${\bf k}_{\rm out}$ for an input from ${\bf k}_{\rm in}$ is given by the Born approximation as proportional to $\tilde{\eta}({\bf q})$, namely the ${\bf q} = {\bf k}_{\rm out} - {\bf k}_{\rm in}$ component of the 3D Fourier transform of the sample's permittivity contrast profile $\eta({\bf r})$~\cite{1969_Wolf_OC, 2017_Jin_JOSAB_review, 2018_Park_nphoton_review, fercher2003oct}.
Such single-scattering contributions add up in phase in Eq.~\eqref{eq: psi_smt} to form the image, similar to an inverse Fourier transform from $\tilde{\eta}({\bf q})$ to $\eta({\bf r})$.
The multiple-scattering contributions do not add up in phase as they differ strongly between different $\bf k_{\rm out}$, $\bf k_{\rm in}$, and $\omega$. In particular, coherently summing over $N$ inputs, outputs, and frequencies raises the single scattering signals $N$ times while the multiple scattering signals only go up by $\sqrt{N}$ times due to their incoherence \cite{kang2015imaging}. Therefore, the triple summations over $\bf k_{\rm out}$, $\bf k_{\rm in}$, and $\omega$ boost the single-to-multiple-scattering ratio in $\psi_{\rm SMT}({\bf r})$. From this reasoning, in theory, as long as there are enough inputs, outputs, and frequencies, multiple scattering can be suppressed below single scattering, leading to deeper imaging depth. }

Next, {we perform spatiotemporal corrections in the virtual experiments to reverse aberrations by computationally engineering the phases of the incident and scattered wavefronts}. The chromatic aberrations in the optical elements of the system and the frequency dependence of the refractive index in the sample create a frequency-dependent phase that misaligns and broadens the temporal gate.
We compensate for such dispersion using a spectral phase $\theta(\omega)$ acting as a virtual pulse shaper 
(Fig.~\ref{fig1:smt}e), like in spectral-domain OCT~\cite{wojtkowski2004ultrahigh,2016_Hillmann_srep}.
The refractive index mismatch between the sample medium ({\it e.g.}, biological tissue), the far field ({\it e.g.}, air), and the coverslip (if there is one) refracts the rays, degrades the input and output spatial gates~\cite{1993_Hell}, and misaligns the spatial gates and the temporal gate. 
By using the appropriate propagation phase shift 
$(\bf k_{\rm out}-\bf k_{\rm in})\cdot{\bf r}$
in each medium and the appropriate arrival time, we restore an ideal spatiotemporal focus at all depths even in the presence of refraction, effectively creating a virtual dry objective lens that reverses both the spatial and the chromatic aberrations arising from the index mismatch (Fig.~\ref{fig1:smt}f) without liquid immersion (Supplementary Sect.~III\,C).
{Importantly, we also digitally introduce
angle-dependent phase profiles $\phi_{\rm in}(\bf k_{\rm in})$ and $\phi_{\rm out}(\bf k_{\rm out})$ that act as two virtual spatial light modulators (SLMs), one for the incident wave and one for the outgoing wave (Fig.~\ref{fig1:smt}g), to compensate for aberrations. These two virtual SLMs can take complicated phase profiles and, thus are able to perform extreme cases of adaptive optics where the corrected sample-induced aberrations are of high-order and fast-varying.}
This yields SMT with corrections,
\begin{equation}
\begin{aligned}
I_{\rm SMT}(\bf{r})\it=&\left| 
\sum_{\omega} e^{i\theta(\omega)}
\sum_{\bf k_{\rm out}}
        e^{i\bf k_{\rm out}\cdot\bf{r} + \it i\phi_{\rm out}(\bf k_{\rm out})} \right. \\
&\,\, \left. 
\sum_{\bf k_{\rm in}} e^{-i\bf k_{\rm in}\cdot\bf{r} + \it i\phi_{\rm in}(\bf k_{\rm in})}
S(\bf k_{\rm out}, \bf k_{\rm in},\omega) \right|^2.
\label{eq:smt_corrected}
\end{aligned}
\end{equation}
By measuring the reflection from a mirror, we remove the dispersion and aberrations in the input path of the optical system (Supplementary Sect.~III\,B).

\subsection{Wavefront Correction \label{sec: sub_opt}}
\begin{figure*}
    \includegraphics[width = 0.88\textwidth]{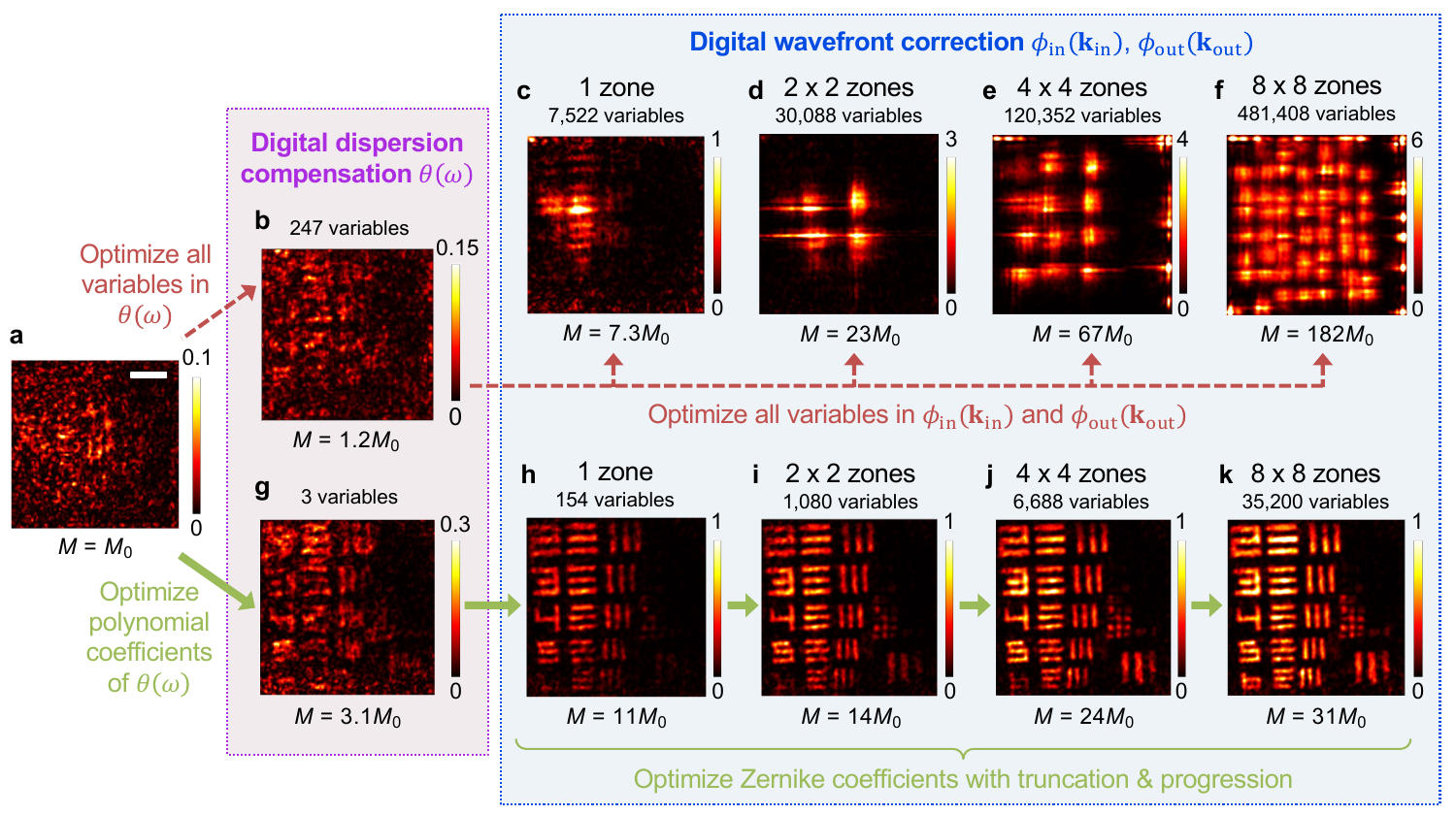}
    \vspace{-6pt}
    \caption{\textbf{SMT digital dispersion compensation and wavefront corrections.}
    SMT finds the digital corrections $\theta(\omega)$, $\phi_{\rm in}(\bf k_{\rm in})$, and $\phi_{\rm out}(\bf k_{\rm out})$ by optimizing an image quality metric $M$, Eq.~\eqref{eq:opt}.
    Different spatial zones of the image use different $\phi_{\rm in/out}$.
    Each panel shows the SMT image $I_{\rm SMT}({\bf r})$ of the USAF-target-under-tissue sample of Fig.~\ref{fig:image2d} in this process, at the depth where $M$ is maximized.
    A direct optimization (red dashed arrows, {\bf b}--{\bf f}) leads to overfitting and gets trapped in poor local optima.
    Through regularization and progression strategies described in the text (green solid arrows, {\bf g}--{\bf k}), SMT finds digital corrections that restore the spatiotemporal focus and enable high-resolution imaging.
    Note that the image before corrections in ({\bf a}) already incorporates triple (temporal, input spatial, and output spatial) gating, index-mismatch correction, and removal of the dispersion and aberrations in the input path of the optical system.
    Scale bar: 10 \textmu m.
    }
    \label{fig:optimization} 
\end{figure*}

Next, we look for suitable corrections $\theta(\omega)$, $\phi_{\rm in}(\bf k_{\rm in})$, and $\phi_{\rm out}(\bf k_{\rm out})$. {Enhancing the focus at $\mathbf{r}$ will enhance the scattering intensity from a target there, so $I_{\rm SMT}(\bf{r})$ acts as an intrinsic virtual guidestar at $\mathbf{r}$ (Fig.~\ref{fig1:smt}c).
Therefore, we maximize an image quality metric $M$,}
\vspace{-2pt}
\begin{equation}
\label{eq:opt}
\hspace{-4pt}
\{\theta, \phi_{\rm in}, \phi_{\rm out}\} = 
\argmax_{\theta, \phi_{\rm in}, \phi_{\rm out}} M, \,\,\,
M = \sum_{\textbf{r}} I_{\rm SMT}(\textbf{r}) \ln \frac{I_{\rm SMT}(\textbf{r})}{I_0}
\end{equation}
with $\sum_{\textbf{r}}$ being a volumetric summation and $I_0$ being a constant; the logarithm factor promotes image sharpness~\cite{1974_Muller_JOSA}.
Similar image metrics have been used in single-path computational AO~\cite{2016_Hillmann_srep} and single-path isoplanatic wavefront shaping with an SLM~\cite{yeminy2021guidestar, 2015_Horstmeyer_nphoton_review};
here we adopt it for digital double-path spatiotemporal corrections.
We derive the gradient of $M$ with respect to the parameters and use a quasi-Newton algorithm, the low-storage Broyden--Fletcher--Goldfarb--Shanno (L-BFGS) method~\cite{1989_Liu_MP}, in the NLopt library~\cite{NLopt} to maximize $M$.
Since a wavefront correction $\phi_{\rm in}(\bf k_{\rm in})$ or $\phi_{\rm out}(\bf k_{\rm out})$ only remains optimal within an isoplanatic patch, ~\cite{2021_Hampson_NRMP_review,2015_Judkewitz_nphys, 2017_Osnabrugge_Optica} 
we divide the space into zones (both axially and laterally) and use different $\phi_{\rm in/out}$ for different zones.

A direct optimization, however, performs poorly when the scattering is sufficiently strong that signals from the target are initially buried under the speckled multiple-scattering background (Fig.~\ref{fig:optimization}a--f).
This happens because (1) the problem is high-dimensional and nonconvex with numerous poor local optima, and (2) the large number of parameters in $\theta(\omega)$, $\phi_{\rm in}(\bf k_{\rm in})$, and $\phi_{\rm out}(\bf k_{\rm out})$ in the many zones can lead to overfitting,
where the optimization inadvertently enhances unintended contributions like the speckled background instead of signals from a target at the intended position ${\bf r}$, (3) {the strong quasi-random multiple scattering signals lead to phase errors in the wavefront correction.}
{To mitigate these issues, we employ five strategies to 
guide the search, avoiding the poor local optima, overfitting, and the phase errors caused by multiple scattering (Supplementary Sect.~III\,E--F):
(i) Regularize $\theta(\omega)$ to a third-order polynomial~\cite{wojtkowski2004ultrahigh} to avoid overfitting. 
{(ii) Regularize $\phi_{\rm in}(\bf k_{\rm in})$ and $\phi_{\rm out}(\bf k_{\rm out})$ each to a Zernike polynomial~\cite{2021_Hampson_NRMP_review, 2007_Debarre} with a controlled number of terms to avoid overfitting and errors by multiple scattering.} 
{(iii) First optimize $\phi_{\rm in/out}$ over the full volume and then progressively optimize over smaller zones while using the previous $\phi_{\rm in/out}$ as the initial guess to efficiently exploit the spatial correlation of sample-induced aberrations, avoiding suboptimal local optima.}
(iv) Truncate the scattering matrix to within the respective spatial zone to avoid overfitting.
(v) Increase the number of Zernike terms as we progress to smaller zones, since the low-order terms correct for slowly varying aberrations with a large isoplanatic volume, while the high-order terms correct for the fast varying aberrations with a small isoplanatic volume.}
With these strategies, we find corrections $\theta(\omega)$, $\phi_{\rm in}(\bf k_{\rm in})$, and $\phi_{\rm out}(\bf k_{\rm out})$ that clearly reveal the target (Fig.~\ref{fig:optimization}g--k) even though {the initial image was overwhelmed by speckles due to strong multiple scattering} despite triple gating (Fig.~\ref{fig:optimization}a).

{While triple gating by itself is not sufficient, it is a necessary ingredient in the optimization because it
promotes the target-to-speckle ratio.
Without the summations
over $\bf k_{\rm out}$, $\bf k_{\rm in}$, and $\omega$
in Eq.~\eqref{eq:smt_corrected}, the optimization may maximize the background speckles instead when scattering is strong.
Since all targets within an isoplanatic volume share the same optimal wavefront while the speckled backgrounds do not, the volumetric summation over space ${\bf r}$
in Eq.~\eqref{eq:opt} also promotes the target-to-speckle weight in the gradient of the optimization.
Supplementary Fig.~10 shows that without one of these quadruple summations in $\{{\bf r}, \bf k_{\rm out}, \bf k_{\rm in}, \omega\}$, 
the optimization landscape may not reflect the target contribution and may exhibit many local optima associated with the multiple-scattering background.}
{This suggests that optimizing over a spatial region where the wavefront distortion remains approximately constant, that is, within an isoplanatic area, is important to improve the optimization robustness and that the optimization will be less affected by poor local optima if the isoplanatic patch is large.
A notable example is the recently proposed image-guided computational holographic wavefront shaping method by Haim et al.~\cite{2024_Haim}, which successfully demonstrated detection-path correction across a large number of scattered modes using iterative wavefront optimization.
While this work achieved high-quality imaging of targets located at a distance behind the complex medium, many real-world applications involve more challenging conditions, such as targets embedded within volumetric scattering media where both the illumination and detection paths are affected by strong, spatially varying distortions. These conditions result in much smaller isoplanatic volumes and stronger multiple scattering, making direct optimization prone to overfitting or convergence to poor local optima and speckle-dominated solutions. SMT addresses such challenges by conducting double-path corrections along with regularization and coarse-to-fine optimization strategies.
}

CLASS~\cite{kang2017high,yoon2020laser,2023_Kwon_ncomms,2023_Li_OE} (which is equivalent to the distortion matrix approach~\cite{2024_Najar_NatComm,2024_Balondrade_NatPhoton,2023_Bureau_ncomms}) and VRM~\cite{2023_Lee_ncomms} are based on the correlations of the single-scattering signal;
there were not formulated as a maximization problem like Eq.~\eqref{eq:opt}.
However, our analysis reveals that they in fact implicitly maximize some metrics (Supplementary Sect.~IV\,D).
But these implicit maximizations were not designed to avoid poor local optima or overfitting, which are important (Fig.~\ref{fig:optimization}).
Also, the dispersion compensation phase for each wavelength of VRM~\cite{2023_Lee_ncomms} is found by correlating the reflected fields of each input at that wavelength and the central wavelength, which are without a confocal gate, a temporal gate, and a volumetric spatial summation, all of which are important (Supplementary Fig.~10) to reduce the error caused by multiple scattering; the other methods, meanwhile, do not compensate for dispersion.
These are part of the reasons the previous matrix-based methods fail when multiple scattering is strong enough to cause errors in the aberration correction.

\begin{figure}
    \includegraphics[scale = 0.88]{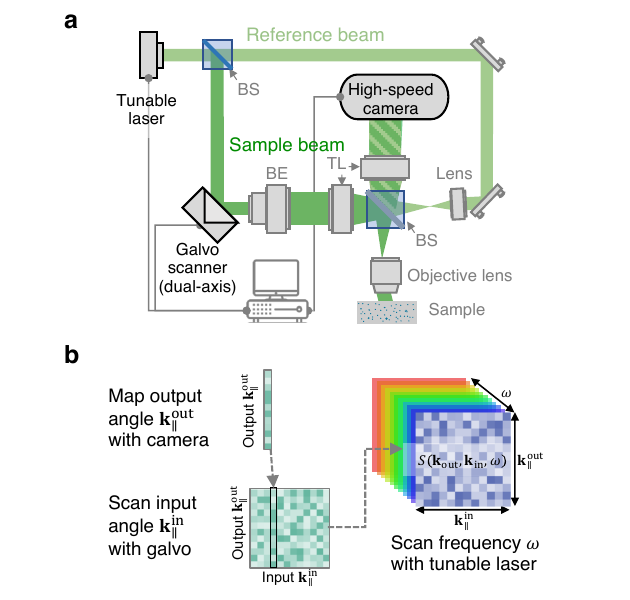}
    \caption{\textbf{Measurement of the hyperspectral reflection matrix.} 
    \textbf{a}, We use off-axis holography to measure the phase and amplitude of fields scattered by the sample.  
    BS: beam splitter; BE: beam expander; TL: tube lens.
    See Supplementary Fig.~1 for a detailed schematic.
    \textbf{b}, 
    Construction of the data cube by mapping the output angles with the camera, scanning the input angle with the galvo, and scanning the frequency with the tunable laser.
    }
    \label{fig:setup} 
\vspace{-4pt}
\end{figure}

\vspace{-6pt}
\section{Experimental demonstration}
\vspace{-3pt}

We use off-axis holography~\cite{2011_Verrier_AO} to measure the scattering matrix in reflection mode, for which
$S({\bf k}_{\rm out}, {\bf k}_{\rm in},\omega)$ reduces to the reflection matrix $R({\bf k}_{\rm out}, {\bf k}_{\rm in},\omega)$ (Fig.~\ref{fig:setup}).
A CMOS camera (Photron Fastcam Nova S6) captures 64,000 columns of the reflection matrix per second, a dual-axis galvo scanner (ScannerMAX Saturn 5B) scans the incident angle ${\bf k}_{\rm in}$, and a tunable laser (M Squared SolsTiS 1600) scans the frequency $\omega$. 
The data acquisition here is two to three orders of magnitude faster than previous measurements of broadband scattering matrices~\cite{2016_Mounaix_PRL,2019_Xiong_ncomms,2023_Lee_ncomms}.
We use a dry objective lens (Mitutoyo M Plan Apo NIR 100X, NA $=$ 0.5).
The power onto the sample ranges from 0.02~mW to 0.2~mW.
We measure around 250 wavelengths $\lambda$ from 740~nm to 940~nm, 2,900 outgoing angles, and 3,900 incident angles within the NA, over a $51\times51$ \textmu m$^2$ area.
The detection sensitivity, currently limited by the residual reflection from the objective lens, is 90 dB.
See Supplementary Sects.~I--II for details.

\vspace{-14pt}
\subsection{Planar Imaging}
\vspace{-8pt}
\begin{figure*}
    \includegraphics[scale = 0.68]{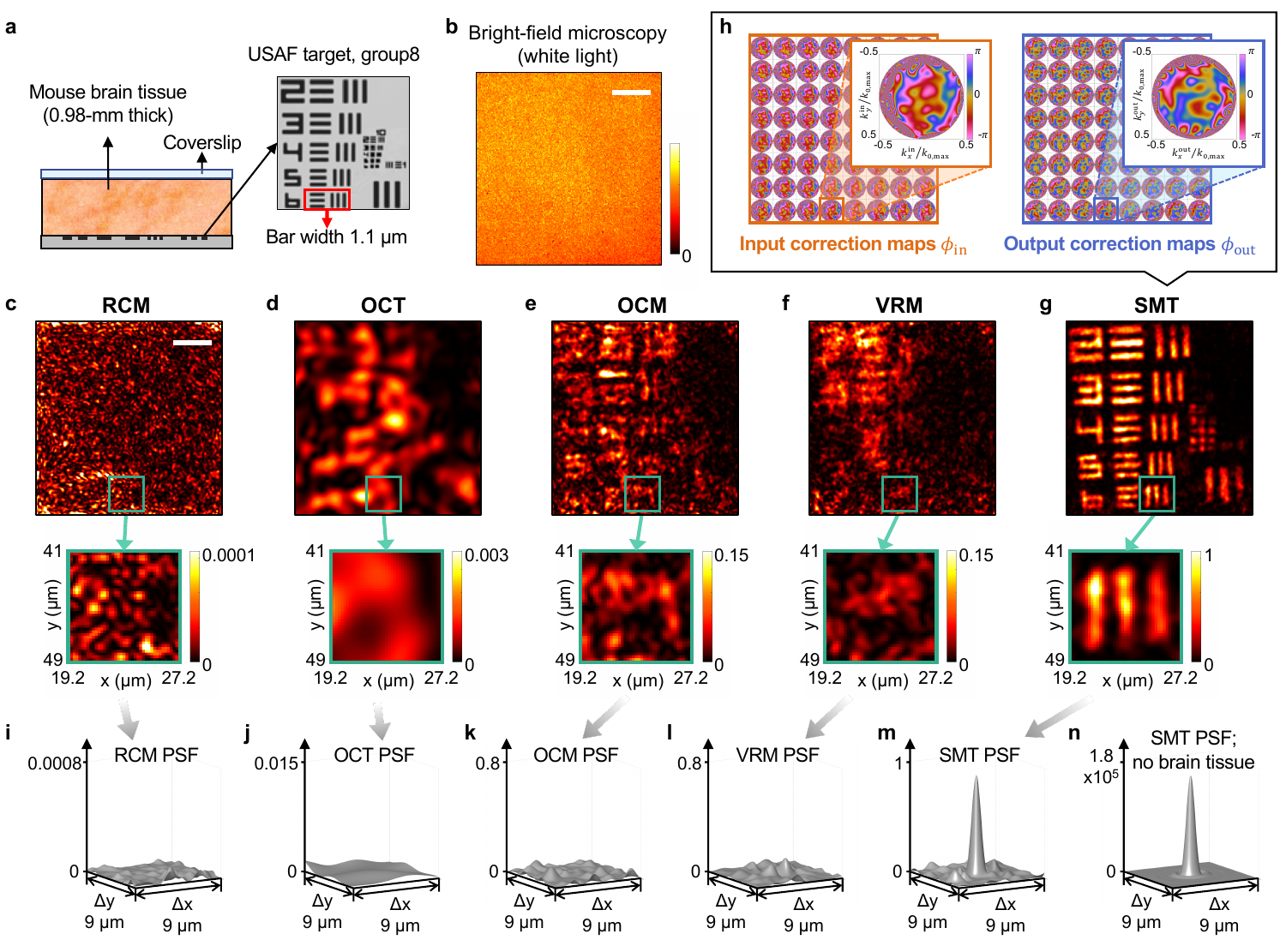}
    \vspace{-8pt}
    \caption{\textbf{Noninvasive imaging through thick tissue.}
    \textbf{a,} Schematic of the sample---a USAF resolution target underneath 0.98 mm of mouse brain tissue---and a scanning electron microscope image of the USAF target before covered by the tissue. 
    \textbf{b,} A standard bright-field microscope image of the sample (with white-light illumination).
    \textbf{c--f,} Reflectance confocal microscopy (RCM), optical coherence tomography (OCT), optical coherence microscopy (OCM), and volumetric reflection-matrix microscopy (VRM) images at the USAF target plane, synthesized from the measured hyperspectral reflection matrix. 
    \textbf{g,} SMT image, $I_{\rm SMT}({\bf r})$, from Eqs.~\eqref{eq:smt_corrected}--\eqref{eq:opt}. 
    Each pair of full view and zoom-in uses the same colorbar, and
    all images share the same normalization, with scales indicated on the colorbars.
    Scale bar in \textbf{b--c}: 10 \textmu m. 
    \textbf{h,} The wavefront correction phase maps for the $8\times8$ zones in SMT. 
    \textbf{i--n,} Corresponding point spread function PSF$({\bf r})$ of the sample ({\bf i--m}) and of a mirror in air ({\bf n}), centered at $(x_{\rm in}, y_{\rm in}) = (23.4, 45.0)$ \textmu m.
    }
    \label{fig:image2d} 
\end{figure*}
{We first image the 8th group (whose 6th element has a bar width and separation of 1.1 \textmu m) of a 1951 USAF resolution target underneath a 0.98-mm-thick tissue slice from the cerebral cortex of a mouse brain (Fig.~\ref{fig:image2d}a)}. 
A standard bright-field microscope image (with incoherent white-light illumination) shows no feature (Fig.~\ref{fig:image2d}b).

\begin{figure*}
    \includegraphics[scale = 0.55] {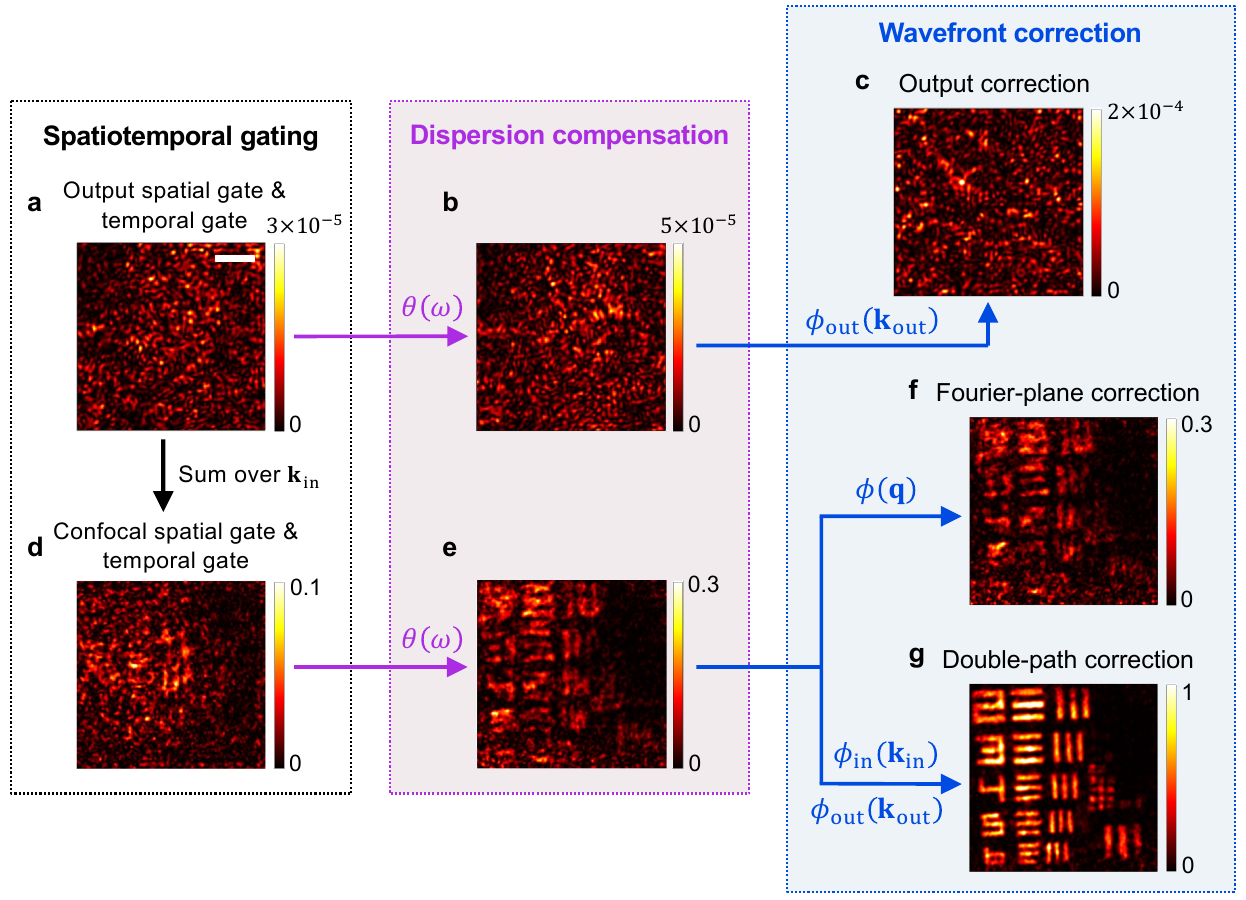}
    \vspace{-6pt}
    \caption{
    \textbf{Role of triple gating and double-path wavefront correction.}
    \textbf{a--c}, Reconstructed image of the USAF target under tissue following the same procedure as SMT but without input spatial gating.
    \textbf{d--f}, Reconstructed image with triple gating but with digital aberration correction only in the reciprocal space {\bf q} of the image.
    \textbf{g}, SMT image with triple gating and double-path wavefront correction.
    Scale bar: 10 \textmu m.
    }
    \vspace{-4pt}
    \label{fig:double_path} 
\end{figure*}

Before considering SMT, we first use the measured hyperspectral reflection matrix to conduct virtual experiments that mimic existing reflection-based imaging methods (Supplementary Sect.~IV):
restricting the frequency summation of Eq.~\eqref{eq:smt_corrected} to one frequency yields a synthetic RCM image without time gating (Fig.~\ref{fig:image2d}c);
restricting the angular summations of Eq.~\eqref{eq:smt_corrected} to small angles (we use NA $=$ 0.1 here) and fixing the depth of the spatial focus yields a synthetic OCT image (Fig.~\ref{fig:image2d}d);
for OCM (Fig.~\ref{fig:image2d}e), we sum over all frequencies and all angles.
The depth of the focal plane is chosen to maximize the total signal.
All of these images include corrections for the input aberrations of the optical system.
For OCT and OCM, we also perform digital dispersion compensation (following the same steps as in SMT). 
{Despite the confocal spatial gate, temporal gate, and corrections, none of these methods reveals any group-8 element due to the overwhelming scattering from the tissue.}
To compare to the most advanced matrix-based method, we also implement VRM~\cite{2023_Lee_ncomms} (Supplementary Sect.~IV\,D), which performs (Fig.~\ref{fig:image2d}f) worse than SMT without the regularization and progression strategies (Fig.~\ref{fig:optimization}c) for reasons explained earlier. 
{In addition, we also benchmarked against the eigenchannel-based approach \cite{badon2016smart,Yang2020extend,cao2022} with the details and results in Supplementary Sect.~IV\,E. In this approach, the scattering matrix is decomposed into singular values and vectors corresponding to distinct optical channels, and then the smaller singular values that correspond to multiple scattering eigenchannels are set to zeros to suppress multiple scattering. 
While this method offers better computational efficiency through its single-step decomposition process, our comparative analysis reveals its limitations in imaging performance. Although it reduces multiple scattering effects in certain regions, it fails in most of the field of view due to the lack of an effective wavefront correction.}

In comparison, the SMT image resolves the 8th group of the USAF target with near perfection down to the smallest 6th element (Fig.~\ref{fig:image2d}g).
Here, the progression goes down to 8$\times$8 zones, with up to 275 Zernike polynomials (22 radial orders) per zone in both $\phi_{\rm in}$ and $\phi_{\rm out}$. 
The optimized correction patterns $\phi_{\rm in}({\bf k}^{\rm in}_{\parallel})$ and $\phi_{\rm out}({\bf k}^{\rm out}_{\parallel})$ are shown in Fig.~\ref{fig:image2d}h. Here, $\phi_{\rm out}({\bf k}_{\parallel}) \neq \phi_{\rm in}(-{\bf k}_{\parallel})$ because the former includes the optical-system aberration.
Notably, SMT achieves good reconstructions even when the number of frequencies or incident angles is further reduced by over an order of magnitude, including when the image remains fully speckled after dispersion compensation (Supplementary Sect.~V). The total number of wavefront correction variables, $2 \times 275 \times 8^2 = 35,200$, is large. Supplementary Fig.~13 shows that all of these terms work together to raise the image metric; there is no redundancy.
The wavefront corrections vary faster at larger angles due to the use of Zernike polynomials as wavefront regularizers, which causes larger momenta to have larger correction phases and become fast-varying after phase wrapping. {Future works will explore various strategies to correct for aberrations at large angles more effectively, such as defining the Zernike polynomials with a larger pupil diameter, alleviating this fast variation.}

The phase-resolved interferometric data in the synthetic OCT and OCM can be used for dispersion compensation and digital aberration correction via computational AO~\cite{adie2012computational,2015_Shemonski_nphoton,2021_Liu_BOE,2016_Hillmann_srep}.
However, the correction only applies to the phase image itself; it cannot apply separately to the incident path and the return path since the reflection matrix is not available in OCT/OCM.
Fig.~\ref{fig:double_path}a--c shows correction on an OCM image without input spatial gate (similar to Ref.~\citen{2016_Hillmann_srep}), which fails here given the lack of triple gating.
Fig.~\ref{fig:double_path}d--f shows correction in the reciprocal space of a confocal OCM image (mimicking Refs.~\citen{adie2012computational,2015_Shemonski_nphoton, 2021_Liu_BOE}; see Supplementary Sect.~IV\,C for details), which yields negligible enhancements because most of the wavefront distortions here are high-order and can only be corrected in a double-path configuration~\cite{2021_Liu_BOE}.
Unklike computational AO, SMT enables triple gating and double-path wavefront correction, which is necessary to overcome the strong scattering here (Fig.~\ref{fig:double_path}g).

To quantify the imaging performance, we obtain the point spread function (PSF) by evaluating Eq.~\eqref{eq:smt_corrected} with a variable output position ${\bf r} = {\bf r}_{\rm in} + \Delta {\bf r}$ given a fixed input position ${\bf r}_{\rm in}$ on the sixth element: ${\rm PSF}({\bf r}) = I({\bf r}, {\bf r}_{\rm in})$; see Supplementary Sect.~VI.
The PSFs of RCM, OCT, OCM, and VRM (Fig.~\ref{fig:image2d}i--l) have no discernible peak near $\Delta{\bf r} = 0$, consistent with their complete failure to image here. 
The speckled OCM PSF averages to be 70~dB below the peak PSF of a mirror without the brain tissue (Fig.~\ref{fig:image2d}n), so {the signal (which is buried beneath the speckled background and not visible here) has been reduced by at least ten-million-fold due to multiple scattering.}
The PSF of SMT (Fig.~\ref{fig:image2d}m) exhibits a sharp peak at $\Delta{\bf r} = 0$ with 7-times the height of the tallest speckle in the background, showing SMT has not reached its depth limit (where the signal strength equals the background strength) yet.
The SMT peak's full width at half maximum (FWHM) is 1.08 \textmu m, close to the mirror PSF's 0.93 \textmu m FWHM, demonstrating diffraction-limited resolution despite the overwhelming multiple scattering.
To our knowledge, the depth-over-FWHM ratio of 910 here is the highest reported in the literature for imaging high-contrast targets inside or behind {\it ex vivo} biological tissue, which scatters about twice as much as {\it in vivo} tissue~\cite{2009_Kobat_OE}.

\vspace{-14pt}
\subsection{Volumetric Imaging}
\vspace{-8pt}

We next perform 3D tomography of a
dense colloid consisting of high-index titanium dioxide (TiO$_2$, refractive index 2.5) nanoparticles dispersed in polydimethylsiloxane (PDMS, refractive index 1.4). The nanoparticles (Sigma-Aldrich, 914320) have a typical diameter of 500~nm, and we use Mie theory to estimate the scattering and transport mean free paths to be $\ell_{\rm sca} = 0.19$~mm and $\ell_{\rm tr} = 0.47$~mm at $\lambda = 840$ nm (Supplementary Sect.~VII).
{We measure the reflection matrix over a $51\times51$ \textmu m$^2$ field of view with the reference plane at depth $z_0 = 1.525$ mm.} Using the reflection matrix, we reconstruct an SMT image over a $50\times50\times110$~\textmu m$^3$ volume inside the colloid by dividing the depth of field (DOF) $\Delta z = 110$~\textmu m  into 16 sub-volumes vertically. 

\begin{figure*}
    \includegraphics[scale = 0.74] {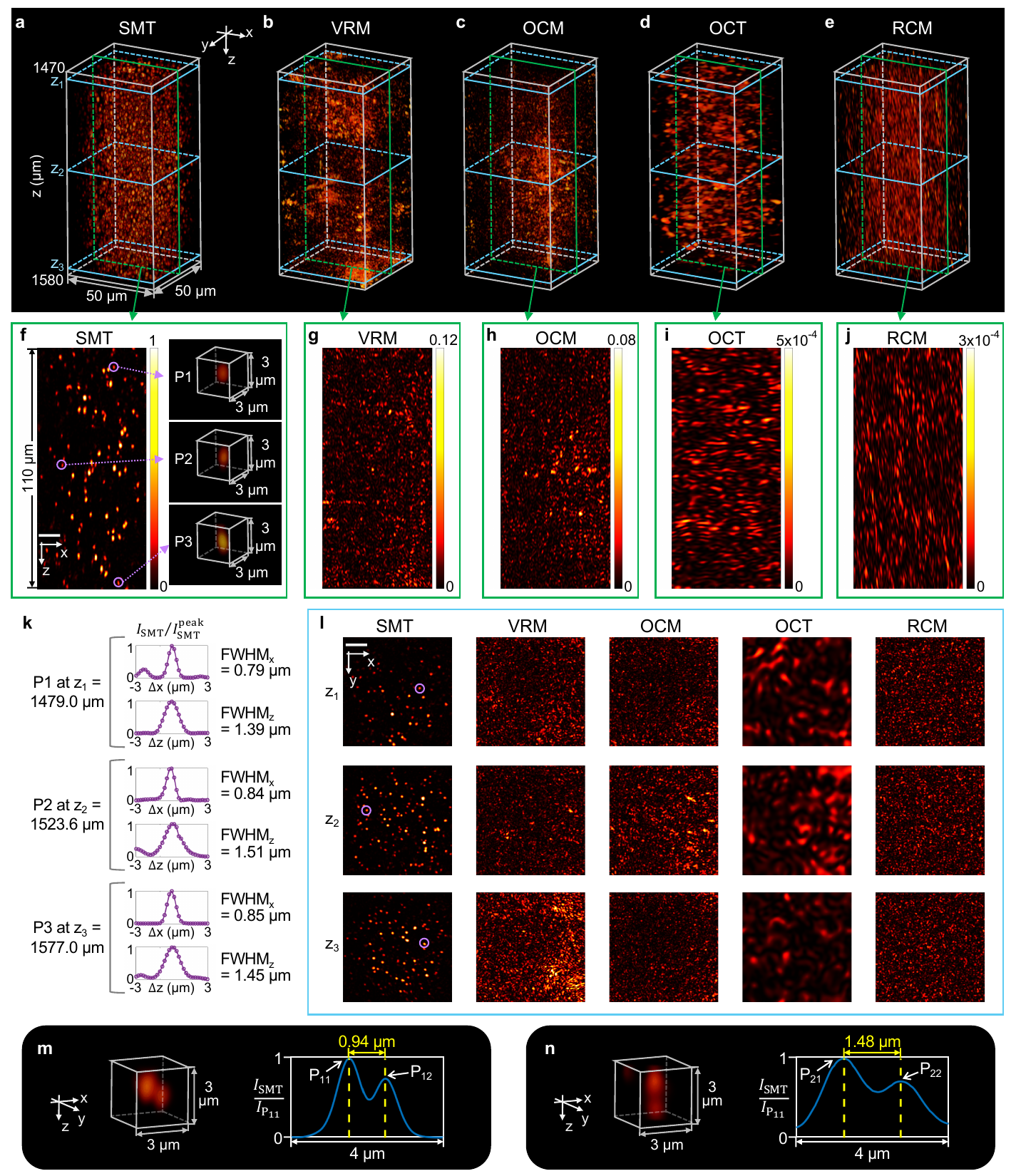}
    \caption{\textbf{Volumetric imaging inside a dense colloid.}
    The sample consists of 500-nm-diameter TiO$_2$ nanoparticles dispersed in PDMS, with an estimated transport mean free path of 0.47 mm.
    \textbf{a--e,} 
    SMT, VRM, OCM, OCT, and RCM images built from the measured hyperspectral reflection matrix. 
    \textbf{f--j,}
    A longitudinal slice of the images at $y = 23.2$ \textmu m and close-up views of three particles at different depths in the SMT image.
    \textbf{k,}
    Cross sections of the three particles;
    $\Delta {\bf r} = {\bf r} - {\bf r}_{\rm peak}$.
    \textbf{l,}
    Transverse slices at the depths of the three particles.
    \textbf{m,n,} SMT images of particles separated horizontally (\textbf{m}) and vertically (\textbf{n}), with center-to-center cross sections. 
    ${\bf r}_{\rm P_{11}} = (46.39, 31.29, 1487.31)$ \textmu m,
    ${\bf r}_{\rm P_{12}} = (46.90, 31.96, 1487.74)$ \textmu m,
    ${\bf r}_{\rm P_{21}} = (44.04, 39.12, 1569.56)$ \textmu m, and
    ${\bf r}_{\rm P_{22}} = (44.18, 38.92, 1571.02)$ \textmu m.
    Scale bars in \textbf{f} and \textbf{l}: 10 \textmu m.
    All images share the same normalization. Volumetric images and 2D slices use the same colorbar.
    }
    \label{fig:image3d} 
\end{figure*}

SMT creates a detailed 3D image of all the nanoparticles in this volume (Fig.~\ref{fig:image3d}a), with
zoom-ins and cross sections of individual particles shown in Fig.~\ref{fig:image3d}f,k,l.
A statistical analysis of these particles finds the lateral FWHM to vary from 0.7 \textmu m to 1.0 \textmu m within the $110$~\textmu m DOF and the axial FWHM very close to the theoretical estimate of 1.42 \textmu m given the spectral bandwidth of $\Delta \lambda = 187$~nm here (Supplementary Sect.~VIII).
Fig.~\ref{fig:image3d}m,n shows examples of SMT images of nearby particle pairs, with one pair separated horizontally by 0.94 \textmu m and one pair separated vertically by 1.48 \textmu m. 
The SMT DOF covers approximately the volume of overlap among the input/output beams in the scattering matrix measurement; it grows with the field of view and is not restricted by the Rayleigh range $z_{\rm R}$ (Supplementary Sect.~VIII\,B).
Here, $z_{\rm R} = n_{\rm sam} \lambda_0 / (\pi {\rm NA}^2) \approx 1.5$ \textmu m (with center wavelength $\lambda_0 = 840$ nm and $n_{\rm sam} = 1.4$ for PDMS),
while the SMT DOF is 73 times larger. 

We also construct VRM (generalized from isolated fixed-depth slices~\cite{2023_Lee_ncomms} to volumetric and vertical slices) and synthetic OCM, OCT, and RCM images (Fig.~\ref{fig:image3d}b--e) for comparison, with
longitudinal and transverse slices shown in Fig.~\ref{fig:image3d}g--j,l.
For OCM and OCT, the depth of the spatial gate is fixed on one focal plane at $z_0=1.525$~mm when the temporal gate is scanned in $z$ (Supplementary Sect.~IV\,B).
{All of these methods fail and cannot identify any of the nanoparticles at these depths due to overwhelming multiple scattering. The initial SMT image is also dominated by speckles (Supplementary Sect. III\,F.7) but eventually reveals nanoparticles with high contrast and resolution thanks to corrections.}.
The VRM image exhibits bright artifacts at several constant-$z$ slices because its dispersion compensation maximizes the intensity at those isolated slices (Supplementary Sect.~IV\,D).

\vspace{-14pt}
\section{Discussion}
\vspace{-8pt}

SMT provides an intuitive and versatile framework for noninvasive label-free computational imaging deep inside scattering media.
By formulating the wavefront correction and image reconstruction into a nonconvex optimization problem and regularizing the wavefronts, SMT sets a framework for future matrix-based imaging works to build upon, as the optimization algorithm, the optimized image quality metrics, and the number of optimized wavefront modes, can be flexibly chosen depending on each particular sample. 
{Although only the $ex~vivo$ mouse brain and dense nanoparticle colloid are used in this paper to primarily demonstrate the capabilities and potentials of SMT, future works will deal with different types of biological tissues such as bone or muscle and non-biological scattering media like white paint or white papers as we work to implement SMT in various practical applications such as medical imaging or material inspection.}

{Compared to existing state-of-the-art matrix-based imaging techniques \cite{yoon2020laser,2023_Lee_ncomms,2024_Balondrade_NatPhoton}, SMT differs in several key aspects: the quadruple summation in image quality metric greatly boosts target-to-speckle ratio; the coarse-to-fine progressive multiscale correction further mitigates local minima caused by strong multiple scattering background noise; and the Zernike-regularized wavefront parameterization avoids overfitting to prevent the optimization from strengthening speckles rather than the target signal. These strategies enable SMT to perform more robust correction in highly scattering, inhomogeneous samples, pushing the imaging depth and quality beyond what current reflection matrix-based techniques can achieve.}

{One aspect not considered in this work is the frequency dependence of the wavefront correction.
It is well known that the optimal wavefront for focusing behind scattering media depends on the frequency~\cite{2011_vanBeijnum_OL}.
Future work can explore optimization strategies that incorporate the frequency dependence in $\phi_{\rm in}$ and $\phi_{\rm out}$ while avoiding overfitting and the larger number of poor local optima due to the additional degrees of freedom (Supplementary Sect. III.F.8).}
One may also explore other optimization schemes beyond using an image quality metric.

Validation against the ground truth is nontrivial for volumetric imaging~\cite{2022_Krauze_srep}.
Using the recent ``augmented partial factorization'' (APF) simulation method~\cite{2022_Lin_NCS}, 
we have validated SMT through full-wave numerical simulations~\cite{2023_Wang}. {In addition, numerical methods like APF or Monte Carlo simulations \cite{2022_Jacques} may help estimate how many inputs, outputs, and frequencies are needed to effectively suppress multiple scattering below single scattering. In particular, simulations of phantoms provide information on how many photons are scattered multiple times versus going ballistic \cite{2022_Cua_BOE}, leading to the ratio $\rho$ between single and multiple scattering signals at a particular imaging depth in real samples. Coherently summing over $N$ inputs, outputs, and frequencies increases the ratio between single and multiple scattering signals by $\sqrt{N}$ times \cite{kang2015imaging}. Guided by simulations, the number of inputs, outputs, and frequencies enough to suppress multiple scattering should be larger than $1/\rho^2$.
{Furthermore, the electric field Monte Carlo methods \cite{Xu2004,Wang2012gpu,Sawicki2008,Radosevich2012} and full-wave simulations like APF \cite{Safadi2023} can model the propagation of complex electric fields inside scattering media. This could offer valuable physical insights to better understand and characterize multiple scattering signals in terms of field-level properties, such as the electric field amplitude, polarization, and phase shifts, showing great potential to further support and inform the developments of SMT. }}
Numerical simulations of modeled phantoms will therefore be employed in future works to guide the design of SMT measurement setups and to develop advanced multiple scattering suppression strategies.

{To ensure successful practical applications, particularly for real-time and {\it in vivo} imaging, several factors must be addressed, and further improvements can be made in subsequent steps.}
One can interpret RCM, OCT, and OCM as measuring the diagonal elements of the reflection matrix in a spatial basis, while SMT, as a matrix-based imaging approach, utilizes the additional off-diagonal elements to digitally refocus to different depths and to enable double-path wavefront corrections.
The price is having to measure those off-diagonal elements.
{While the use of all matrix elements enables high-quality imaging, it presents scalability challenges for large field-of-view applications, particularly for highly dynamic samples. This is because the number of matrix elements scales quadratically with the number of diffraction limits within the field-of-view.
These additional measurements require faster acquisition, especially in {\it in vivo} imaging, where
motion artifacts make aberration correction and imaging {\it in vivo} a challenge~\cite{2017_Liu_Optica}.
While the virtual imaging and wavefront correction of SMT dispenses with slow SLMs, a coherent synthesis still requires us to complete the measurement of the matrix elements before the scatterer arrangement changes substantially. {In addition, as a computational imaging and aberration correction method, SMT requires accurate phase information. The motion of samples, especially living samples, can create a significant phase instability. To avoid motion artifacts and motion-induced phase instability, the most straightforward way is to expedite the measurement.}
The speckle decorrelation time, primarily limited by the blood flow, is estimated as 5 ms at $\lambda = 633$ nm wavelength 1 mm deep inside the mouse brain where blood is rich~\cite{2017_Qureshi_BOE} but can go beyond 30 seconds for the skull where there is less blood~\cite{2023_Kwon_ncomms}.

{These challenges in scalability and {\it in vivo} measurements open up important, promising directions. 
Of particular interest is investigating the relative significance of different matrix elements, which could enable selective measurements to reduce both acquisition time and computational demands. Specifically, instead of measuring the scattering matrix in the angular basis, measurement can be done in the spatial basis where only the reflection from output locations $\mathbf{r}_{\rm out}$ close to the incident location $\mathbf{r}_{\rm in}$ need to be collected, dramatically reducing the number of matrix elements to measure,~\cite{2024_Najar_NatComm} leading to faster measurement.}
{Also, encouragingly, other ongoing efforts to improve the measurement speed of matrix-based imaging~\cite{2023_Lee_ncomms,2024_Najar_NatComm,2024_Balondrade_NatPhoton}, alongside advancements in hardware development, are making SMT increasingly practical for real-world applications.}
{Currently, our total measurement time for the USAF-target-under-tissue sample is three minutes, with the majority of which was spent on the mechanical acceleration and deceleration of a birefringent filter during a scan-and-stop operation of the tunable laser.}
Scanning the frequency continuously can reduce the measurement time to 5 seconds (Supplementary Sect.~I\,D). 
{One can further accelerate by orders of magnitude by reducing the number of frequencies or incident angles (Supplementary Sect.~V)}.
Advancements in hardware are also playing a key role in improving measurement speed.
{The use of high-speed swept sources can significantly reduce spectral scanning times, while innovations to parallelize the spectral acquisition, such as imaging mapping spectrometers~\cite{2014_Gao_jbio_review}, allow for the single-shot capture of multiple spectral images, further expediting the data collection.
With such accelerations, {\it in vivo} SMT data acquisition is possible even inside blood-rich areas like the brain.}

Beyond accelerating data acquisition, reducing the time spent on computational processing is equally important for practical use, especially for real-time applications.
{For SMT as well as other matrix-based imaging methods, the primary challenges stem from both data transfer speeds and computational intensity.
Although the computations are performed after all data is acquired, so in principle, it does not hinder the prospect of $in~vivo$ imaging, the long processing time is a significant challenge for real-time imaging, where all the data must be transferred and processed in a short amount of time.
The current implementation of SMT employs nonuniform fast Fourier Transform for image reconstruction, achieving competitive processing efficiency compared to conventional matrix multiplication methods common in other matrix-based techniques \cite{2024_Najar_NatComm,2024_Balondrade_NatPhoton,2023_Bureau_ncomms}. However, significant challenges remain for real-time applications.
Currently, the time spent on processing raw data, finding suitable wavefront correction phases, and reconstructing images may range from a few minutes to hours while data transferring between hardware components can take up to 20 minutes (Supplementary Sect.~III.H) due to the large data size.
Potential graphics processing units (GPU) acceleration~\cite{2024_Balondrade_NatPhoton,2023_Bureau_ncomms}, coupled with ongoing developments in data transmission technology and camera interfaces, as well as selective matrix element measurements, suggest promising pathways for enhancing SMT's practical utility, particularly in applications where rapid feedback is crucial.
}

{Another aspect of SMT 
worth exploring is the ability to integrate it with other imaging modalities for practical applications. When measuring the matrix in the angular basis, SMT illuminates the samples with plane waves with varied incident angles and scanned frequencies. This makes SMT a more generalized version of full-field swept-source OCT \cite{2024_Balondrade_NatPhoton}. Alternatively, SMT can also measure the matrix in the spatial basis, where it employs point illuminations like in OCT/OCM but captures the scattering from not only the illuminated locations but also from other surrounding positions. Therefore, SMT can be modified to be used in conjunction with OCT/OCM, sharing the same illumination part. In addition, as SMT provides a powerful aberration correction, its correction maps can be useful to counteract aberrations in fluorescence or non-linear microscopy, where the phase information is usually difficult to obtain, making computational aberration correction difficult. In Supplementary Sect.~IX, we propose some future optical systems where SMT can be integrated into optical-fiber-based swept-source OCT for surgical applications and fluorescence microscopy for biological applications.}

While SMT uses the scattering strength as the contrast, digital staining~\cite{2019_Rivenson_LSA,2023_Chen_nphoton} and dynamic variation~\cite{2020_Scholler_LSA, 2012_Jia} may be used to infer other contrasts. {Inspired by OCT angiography \cite{2012_Jia}, SMT can also image the decorrelation of highly dynamic samples such as blood vessels (Supplementary Sect. X), with the difference being each SMT image has higher quality than each OCT image due to better aberration correction, potentially revealing more sample features.}
The phase information in $\psi_{\rm SMT}$ can help detect sub-nm displacements~\cite{2004_Akkin_OE, 2019_Kim_BOE}. 
One may incorporate polarization gating to select birefringent objects such as directionally oriented tissues. {As SMT is capable of measuring the full vectorial scattering matrices  (Supplementary Sect. XI), future works can explore how samples interact with light of different polarization.}
The hyperspectral scattering matrix can additionally resolve spectral information of the sample, such as the oxygenation of the hemoglobin.


\vspace{12pt}
\noindent {\bf Data availability:}
The datasets generated and analyzed during the current study are available on Zenodo~\cite{SMT_data}.

\vspace{3pt}
\noindent {\bf Code availability:}
The codes used to produce the results of the study are available on GitHub~\cite{SMTgithub}.

\vspace{3pt}
\noindent {\bf Acknowledgments:}
We thank B.~Applegate, S.~Fraser, O.~D.~Miller, and B.~Flanagan for helpful discussions.
This work was supported by the Chan Zuckerberg Initiative, the National Science Foundation CAREER award (ECCS-2146021), and the University of Southern California.

\vspace{3pt}
\noindent {\bf Author contributions:}
Y.Z. and C.W.H. designed the experimental setup. Y.Z. built the setup, prepared the samples, and carried out the measurements with help from M.D. M.D. and Y.Z. developed and carried out the dispersion compensation. M.D. developed and carried out the SMT wavefront optimization with regularization and progression. Z.W. developed fast 3D reconstruction using the non-uniform fast Fourier transform. Y.Z. and Z.W. developed the index-mismatch correction. Y.Z., T.Z., and T.C. developed the control and automation of the instruments. M.D. and Y.Z. implemented the synthesis of RCM, OCT, and OCM. M.D. implemented VRM. Y.Z. performed the 3D visualization and the analyses on phase stability, sensitivity, and resolution. C.W.H. conceived the project and supervised research. C.W.H., Y.Z., and M.D. wrote the manuscript with inputs from the other coauthors. All authors discussed the results.

\vspace{3pt}
\noindent {\bf Competing interests:}
C.W.H, Z.W., Y.Z., and M.D. are inventors of US patent applications 17/972,073 and 63/472,900 titled ``Multi-spectral scattering-matrix tomography'' filed by the University of Southern California.


\nocite{2013_Yu_PRL, wang2016reducing, camera, riidatabase_glass, Daimon:07, Gupta2019, riidatabase, 2023_Fricker_ZernikeMatlab, 2015_Hsu_PRL, 2019_South_OL, 2013_Kumar_OE, theivasanthi2013titanium, mohanty2015fabrication, matzler2002matlab, DeVore:51, correlation, carminati_schotland_2021, duelk2015sleds, trackmate2017, Fiji, Park2023, trackmate2022}

\bibliographystyle{naturemag} 
\bibliography{bib_combined}      

\end{document}